\documentclass[%
 reprint,
showpacs,preprintnumbers,
 amsmath,amssymb,
 aps,
]{revtex4-1}

\usepackage{graphicx}
\usepackage{dcolumn}
\usepackage{bm}

\begin{document}

\title{Photonic forces in the near field of statistically homogeneous fluctuating
sources}

\author{Juan Miguel Au\~{n}\'{o}n }
\author{Manuel Nieto-Vesperinas}
\email{mnieto@icmm.csic.es}

\affiliation{Instituto de Ciencia de Materiales de Madrid, Consejo Superior de
Investigaciones Cient\'{i}ficas\\
 Campus de Cantoblanco, Madrid 28049, Spain}
 
\begin{abstract}
Electromagnetic sources, as e.g. lasers, antennas, diffusers or thermal
sources, produce a wavefield that interacts with objects to transfer
them its momentum. We show that the photonic force exerted on a small
particle in the near field of a planar statistically homogeneous fluctuating
source uniquely depends and acts along the coordinate perpendicular
to its surface. The gradient part of this force is contributed by
only the evanescent components of the emitted field, its sign being
opposite to that of the real part of the particle polarizability.
The non-conservative force part is uniquely due to the propagating
components, being repulsive and constant. Also, the source coherence
length adds a degree of freedom since it largely affects these forces.
The excitation of plasmons in the source surface drastically enhances
the gradient force. Hence, partially coherent wavefields from fluctuating
sources constitute new concepts for particle manipulation at the subwavelength
scale 
\end{abstract}



\maketitle

\section{INTRODUCTION}

Photonic forces are increasingly studied due to their potential in
many disciplines ranging from physics and chemistry to biology \citep{block1989compliance,block1990bead, nieto2004near}.
Of special importance is the manipulation of dipolar particles, understood
as those for which the incident wave excites their first electric
and/or magnetic Mie coefficients \citep{nieto2010optical, nietoJOSA011}. Extensive
studies done on light from quasi-coherent sources show that it exerts
mechanical action on these particles through both their conservative
(gradient) and non-conservative components, allowing the design of
optical tweezers which rely on the former component \citep{block1989compliance,block1990bead,juan2011plasmon}
and their recent extensions to the subwavelength , particularly nanometric
scale \citep{juan2011plasmon,Dholakia}. On the other hand, the scattering,
or radiation pressure, force component which until recently was believed
to push objects \citep{ChaumetOL, Dholakia,nieto2004near,nieto2010optical,Albala},
has recently been designed to exert a pulling action towards the coherent
source, as recently shown by exciting the induced magnetic dipole
or multipoles of the particle \citep{chen2011optical,novitsky2011single},
as well as by an appropriate design of the illuminating wavefield
angular spectrum \citep{sukhov2011negative}.

We report here a new area of study for optical manipulation at the
subwavelength scale, both theoretical and experimental, by partially
coherent fields emanating from fluctuating sources \cite{mandel1995optical, James2, carney1, kim2009momentum}. They completely
change the nature of these forces and convey new behaviours to them.
Interestingly, we find that planar sources, of such a general class
as those that are statistically stationary and homogeneous, produce
gradient forces that may be either attractive or repulsive. In turn,
we demonstrate that these forces are dramatically enhanced as the
coherence length of the source decreases, as well as when surface
plasmons (SPP) are excited on its surface. On the other hand, the
non-conservative part of the force, composed of the radiation pressure
plus spin density of angular momentum of the electric wavevector,
is pushing and constant throughout the emission half-space. In this
way, one can control the tractor or pushing effect of the resulting
force on the particle according to the sign of the real part of its
polarizability \citep{bourret1960coherence,mehta1964coherence,wolf1984fields,carminati1999near,apostol2003spatial,roychowdhury2003effects,carminati2010subwavelength,aunon2011near,saenz2011optical,sukhov2011negative,electrodynamics1998jd,ashkin1970acceleration,ashkin1986observation}.

\section{FLUCTUATING OPTICAL FORCES}

Let us consider a fluctuating source emitting from the plane $z=0$,
(see Fig. 1). We shall assume that the radiated random field is described
by an ensemble which is stationary, then we may work in the space-frequency
domain \citep{mandel1995optical} so that its electric vector is expressed
at frequency $\omega$ as an angular spectrum of plane waves propagating
throughout the half-space $z>0$ \citep{mandel1995optical,nietolibro}:
\begin{equation}
\mathbf{E}(\mathbf{r},\omega)=\int_{-\infty}^{\infty}\mathbf{e}(k\mathbf{s}_{\perp},\omega)e^{ik\mathbf{s}\cdot\mathbf{r}}d^{2}{\bf s}_{\perp},\label{angspectr}
\end{equation}
 where $k=\omega/c$, $c$ being the speed of light in vacuum. The
propagation vector $\mathbf{k}=k\mathbf{s}$ is expressed as $\mathbf{k}=k(\mathbf{s}_{\perp},s_{z})$,
so that $\mathbf{s}_{\perp}=(s_{x},s_{y})$ are the transversal components
of ${\bf s}$ and $s_{z}=\sqrt{1-|\mathbf{s}_{\perp}|^{2}}$, ($|\mathbf{s}_{\perp}|^{2}\leq1$),
for homogeneous or propagating waves, and $s_{z}=i\sqrt{|\mathbf{s}_{\perp}|^{2}-1}$,
($|\mathbf{s}_{\perp}|^{2}>1$) , for evanescent components .

Let a dipolar particle with dynamic electric polarizability $\alpha_{e}$,
be placed in the source vicinity. Being ${\bf p}=\alpha_{e}{\bf E}$
the dipole moment induced in the particle by the ${\bf E}$ field,
the $ith$ Cartesian component ($i=1,2,3)$ of the mean force that
the emitted wavefield exerts on it at frequency $\omega$ is \citep{nieto2010optical,nieto2004near,wong2006gradient}

\begin{eqnarray}
F_{i}\left(\mathbf{r},\omega\right) & = & \frac{1}{2}\Re\left\{ \alpha_{e}\left\langle E_{j}^{*}\partial_{i}E_{j}\right\rangle \right\} \nonumber \\
 & = & \frac{1}{4}\Re\alpha_{e}\partial_{i}\left\langle E_{j}^{*}E_{j}\right\rangle +\frac{1}{2}\Im\alpha_{e}\Im\left\{ \left\langle E_{j}^{*}\partial_{i}E_{j}\right\rangle \right\} \nonumber \\
 & = & F_{i}^{grad}\left(\mathbf{r},\omega\right)+F_{i}^{nc}\left(\mathbf{r},\omega\right),(i,j=1,2,3),\label{force1}
\end{eqnarray}
expressed as the sum of a conservative, or gradient force, $F_{i}^{grad}$
proportional to $\Re\alpha_{e}$ and a non-conservative term $F_{i}^{nc}$
proportional to $\Im\alpha_{e}$. $\Re$ and $\Im$ stand for real
ad imaginary parts, respectively. The symbol $\ast$ denotes complex
conjugate. The angular brackets mean ensemble average. Einstein's
convention of omitting the sum symbol $\sum_{j=1}^{3}$ on the repeated
index $j$ has been used.

\begin{figure}
\begin{centering}
\includegraphics[scale=0.4]{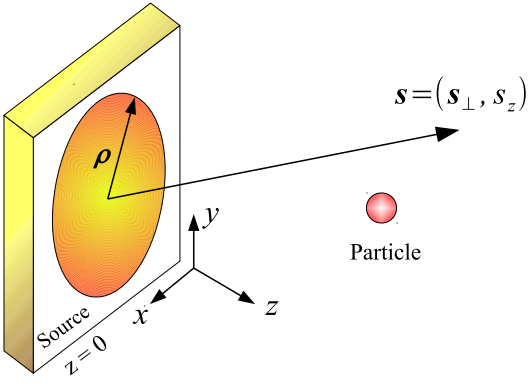} 
\par\end{centering}

\caption{Illustrating the notation}
\end{figure}

On introducing Eq. (\ref{angspectr}) into (\ref{force1}) one obtains
\begin{align}
F_{i}^{grad}\left(\mathbf{r},\omega\right) & =-i\frac{k}{4}\Re\alpha_{e}\iint_{-\infty}^{\infty}\textrm{Tr}\mathcal{A}_{jk}^{(e)}\left(k\mathbf{s}_{\perp},k\mathbf{s'}_{\perp}\omega\right)\nonumber \\
 & \times\left(s_{i}^{*}-s'_{i}\right)e^{-ik\left(\mathbf{s}^{*}-\mathbf{s'}\right)\cdot\mathbf{r}}d^{2}{\bf s}_{\perp}d^{2}{\bf s}'_{\perp},\label{forcegrad}
\end{align}
 
\begin{align}
F_{i}^{nc}\left(\mathbf{r},\omega\right) & =\frac{1}{2}\Im\alpha_{e}\Im\left\{ ik\iint_{-\infty}^{\infty}\textrm{Tr}\mathcal{A}_{jk}^{(e)}\left(k\mathbf{s}_{\perp},k\mathbf{s'}_{\perp}\omega\right)\right.\nonumber \\
 & \times\left.s'_{i}e^{-ik\left(\mathbf{s}^{*}-\mathbf{s'}\right)\cdot\mathbf{r}}d^{2}{\bf s}_{\perp}d^{2}{\bf s}'_{\perp}\right\} ,\label{forcenc}
\end{align}
$(i,j,k=1,2,3)$, $\textrm{Tr}$ denotes the trace of the electric
angular correlation tensor $\mathcal{A}_{jk}^{(e)}\left(k\mathbf{s}_{\perp},k\mathbf{s'}_{\perp},\omega\right)=\left\langle e_{j}^{*}(k\mathbf{s}_{\perp},\omega)e_{k}(k\mathbf{s'}_{\perp},\omega)\right\rangle $.
Notice that since $\left\langle E_{j}^{*}E_{j}\right\rangle $ is
real and non-negative, $F_{i}^{grad}$ given by Eq. (\ref{forcegrad})
which equals $\frac{1}{4}\Re\alpha_{e}\partial_{i}\left\langle E_{j}^{*}E_{j}\right\rangle $
according to Eq. (\ref{force1}), is a real quantity. Eqs. (\ref{forcegrad})
and (\ref{forcenc}) reveal that whereas the gradient force depends
on a weighted sum of the difference vectors $\mathbf{s}^{*}-\mathbf{s'}$
and, as we shall see, it has a negative sign if $\Re\alpha_{e}$ is
positive, thus pulling the particle towards the source, the non-conservative
force associated to $\Im\alpha_{e}$ which is always non-negative,
only depends on the weighted sum of vectors $\mathbf{s}$ and pushes
the particle forward.
\subsection{Statistically homogeneous sources. Gradient and non-conservative  forces}

Let us address the wide variety of \textit{statistically homogeneous
sources}, \citep{bourret1960coherence,mehta1964coherence}. Then their
electric cross-spectral density tensor \citep{mandel1995optical}
${\cal E}_{ij}\left(\mathbf{r}_{1},\mathbf{r}_{2},\omega\right)=\left\langle E_{i}^{*}\left(\boldsymbol{r}_{1}\right)E_{j}\left(\boldsymbol{r}_{2}\right)\right\rangle $
in the source plane $z=0$ is \citep{gbur1999energy} ${\cal E}_{ij}^{(0)}\left(\boldsymbol{\rho}_{1},\boldsymbol{\rho}_{2},\omega\right)={\cal E}_{ij}^{(0)}\left(\boldsymbol{\rho},\omega\right)$,
$\boldsymbol{\rho}=\boldsymbol{\rho}_{2}-\boldsymbol{\rho}_{1}$;
${\bf r}_{\alpha}=(\boldsymbol{\rho}_{\alpha},z_{\alpha})$, ${\alpha}=1,2$.

It is well-known \citep{mandel1995optical} that $\mathcal{A}_{jk}^{(e)}\left(k\mathbf{s}_{\perp},k\mathbf{s'}_{\perp}\omega\right)=k^{2}\mathcal{\tilde{E}}_{jk}\left(k\mathbf{s}_{\perp},k\mathbf{s'}_{\perp}\omega\right)$,
where $\mathcal{\tilde{E}}_{jk}\left(k\mathbf{s}_{\perp},k\mathbf{s'}_{\perp}\omega\right)$
is the four-dimensional inverse Fourier transform of ${\cal E}_{ij}^{(0)}\left(\boldsymbol{\rho}_{1},\boldsymbol{\rho}_{2},\omega\right)$.
In addition, it was proven \citep{wolf1984fields} that for a homogeneous
source the components of the electric angular correlation tensor are
$\delta-$correlated as
\begin{eqnarray}
\mathcal{A}_{jk}^{(e)}\left(k\mathbf{s}_{\perp},k\mathbf{s'}_{\perp},\omega\right)=k^{4}\delta^{\left(2\right)}\left[k\left(\mathbf{s}_{\perp}-\mathbf{s'}_{\perp}\right),\omega\right]\nonumber \\
\times \mathcal{\tilde{E}}_{jk}^{(0)}\left[\frac{k}{2}\left(\mathbf{s}_{\perp}+\mathbf{s'}_{\perp}\right),\omega\right];
\end{eqnarray}
$\delta^{(2)}$ representing the two-dimensional Dirac-delta function.

On introducing the above $\delta$-function expression for $\mathcal{A}_{jk}^{(e)}\left(k\mathbf{s}_{\perp},k\mathbf{s'}_{\perp},\omega\right)$
into Eqs. (\ref{forcegrad}) and (\ref{forcenc}) one straightforwardly
obtains for the gradient force
\begin{eqnarray}
F_{i}^{grad}\left(z,\omega\right) & = & F_{z,ev}^{grad}\left(z,\omega\right)\nonumber \\
 & = & -i\frac{k^{3}}{4}\Re\alpha_{e}\int_{|\mathbf{s}_{\perp}|^{2}>1}\textrm{Tr}\mathcal{\tilde{E}}_{jk}^{(0)}\left(k\mathbf{s}_{\perp},\omega\right)\nonumber \\
 & \times & \left(s_{i}^{*}-s_{i}\right)e^{-2k\sqrt{|\mathbf{s}_{\perp}|^{2}-1}z}d^{2}{\bf s}_{\perp},\label{Fgradhom}
\end{eqnarray}
 The subindex in the integral of Eq. (\ref{Fgradhom}) means that
the integration only extends to the non-radiative region because the
difference vector $\mathbf{s}^{*}-\mathbf{s}$ in Eq. (\ref{forcegrad})
is clearly zero for propagating waves, ($|\mathbf{s}_{\perp}|^{2}\leq1$).\textit{
Therefore the radiative components of the field emitted by statistically
homogeneous sources do not contribute to the gradient force, which
only depends on the evanescent components}, ($|\mathbf{s}_{\perp}|^{2}>1$),
for which $\mathbf{s}^{*}-\mathbf{s}=(0,0,s_{z}^{*}-s_{z})=(0,0,-2i\sqrt{|\mathbf{s}_{\perp}|^{2}-1})$.
\textit{Hence this force only exists in the near field, and depends
on the distance $z$ of the particle to the source, having solely
$z$- component normal to its surface. In addition, this force is
attractive or repulsive depending on the sign of $\Re\alpha_{e}$}.
Small particles with relative permittivity $\epsilon>1$ have $\Re\alpha_{e}>0$
out of resonance and thus $F_{z}^{grad}\left(z,\omega\right)$ will
drag them towards the source. Conversely, near a resonance $\Re\alpha_{e}$
may be negative \citep{nieto2004near}, thus this force being repulsive.
However, further study is required in this latter case, since then
the particle strongly scatterers the field emitted by the source,
and therefore the analysis developed here should not be exact due
to multiple scattering of the radiation between the source and the
particle. Hence it is shown that the gradient force near a statistically
homogeneous source is entirely of non-radiative nature and may work
as a \textit{tractor force} \citep{chen2011optical,novitsky2011single,saenz2011optical,sukhov2011negative}.

Analogously, from Eq. (\ref{forcenc}) one also derives for the non-conservative
force $F_{i}^{nc}$ a dependence on $z$ only: 
\begin{eqnarray}
&&F_{i}^{nc}\left(z,\omega\right)\nonumber\\
&=&F_{i,h}^{nc}\left(z,\omega\right)+F_{i,ev}^{nc}\left(z,\omega\right)\nonumber\\
&=&\frac{k^{3}}{2}\Im\alpha_{e}\Im\left\{ i\int_{\left|\mathbf{s}_{\perp}\right|^{2}\leq 1}\text{Tr}\tilde{{\cal E}}_{jk}^{(0)}\left(k\mathbf{s}_{\perp},\omega\right)s_{i}d^{2}\mathbf{s}_{\perp}\right\}\nonumber \\
&+&\frac{k^{3}}{2}\Im\alpha_{e}\Im\left\{ i\int_{\left|\mathbf{s}_{\perp}\right|^{2}>1}\text{Tr}\tilde{{\cal E}}_{jk}^{(0)}\left(k\mathbf{s}_{\perp},\omega\right)\right.\nonumber\\
&\times &\left.s_{i}e^{-2k\sqrt{\left|\mathbf{s}_{\perp}\right|^{2}-1}z}d^{2}\mathbf{s}_{\perp}\right\},\label{Fnchye}
\end{eqnarray}
 $F_{i,h}^{nc}$ and $F_{i,ev}^{nc}$, denote propagating and evanescent
wave contributions, which correspond to the first and second integral
terms of Eq. (\ref{Fnchye}), respectively. Notice that $F_{i,h}^{nc}>0$
is constant throughout $z>0$ .

Let the source also be statistically isotropic \citep{mandel1995optical}
so that ${\cal E}_{ij}^{(0)}\left(\boldsymbol{\rho}_{1},\boldsymbol{\rho}_{2},\omega\right)={\cal E}_{ij}^{(0)}\left(\rho,\omega\right)$,
where $\rho=|\boldsymbol{\rho}_{1}-\boldsymbol{\rho}_{2}|$. The spatial
coherence function of the field in $z=0$ is \citep{gbur1999energy,WolfCarter1975Lambertian,Setala2003Universality}
$\textrm{Tr}{\cal E}_{ij}^{(0)}\left(\rho,\omega\right)$ and the
\textit{spectral degree of spatial coherence} $\mu^{(0)}\left(\rho,\omega\right)=\textrm{Tr}{\cal E}_{ij}^{(0)}\left(\rho,\omega\right)/S^{(0)}(\omega)$,
where the wavefield spectrum on the source is: $S^{(0)}(\omega)=\textrm{Tr}{\cal E}_{ij}^{(0)}\left(0,\omega\right)$.

To illustrate these results, we shall consider a Gaussian spectral
degree of coherence $\mu^{(0)}\left(\rho,\omega\right)=\textrm{exp}\left[-\rho^{2}/2\sigma^{2}\right]$,
so that taking Fourier inverse one obtains
\begin{eqnarray}
 \textrm{Tr}\mathcal{\tilde{E}}_{jk}^{(0)}\left(k\mathbf{s}_{\perp},\omega\right)&=&S^{(0)}(\omega)\tilde{\mu}^{(0)}(k\mathbf{s}_{\perp},\omega)\nonumber \\
&=&S^{(0)}(\omega)\left(\sigma^{2}/2\pi\right)\textrm{exp}\left[-\left(k\sigma\left|{\bf s}_{\perp}\right|\right)^{2}/2\right],\nonumber\\
&&
\end{eqnarray}
where $\sigma$ is the correlation or coherence length of the source.
On introducing this expression for $\textrm{Tr}\mathcal{\tilde{E}}_{jk}^{(0)}\left(k\mathbf{s}_{\perp},\omega\right)$
into the force equations (\ref{Fgradhom}) and (\ref{Fnchye}), we
obtain that on writing in cylindrical coordinates: $s_{x}=s_{\perp}\cos\phi$,
$s_{y}=s_{\perp}\sin\phi$, and due to the rotational symmetry of
the source, the transversal components of the non-conservative force
are zero, viz. : $F_{x,y}^{nc}\left(z,\omega\right)=0$ since so are
the corresponding integrals of Eq. (\ref{Fnchye}) when one performs
the azimuthal angle $\phi$ integration. Also, since $s_{z}=i\sqrt{|\mathbf{s}_{\perp}|^{2}-1}$
for $|\mathbf{s}_{\perp}|^{2}>1$, the second integral in Eq. (\ref{Fnchye})
is purely imaginary, which implies that $F_{z,ev}^{nc}=0$.

\begin{eqnarray}
&&F_{i}^{nc}(z,\omega)=F_{i,h}^{nc}(z,\omega)\nonumber \\
&=&\frac{k^{3}}{2}\Im\alpha_{e}\Im\left\{ i\int_{\left|\mathbf{s}_{\perp}\right|^{2}\leq1}\textrm{Tr}\mathcal{\tilde{E}}_{jk}^{(0)}\left(k\mathbf{s}_{\perp},\omega\right)s_{i}d^{2}{\bf s}_{\perp}\right\} .
\label{Fncho}
\end{eqnarray}
Thus, while $F_{z,h}^{nc}\left(z,\omega\right)>0$ is constant throughout
$z>0$, as so is the spectrum $S^{(0)}(\omega)$ propagating into
$z>0$ \citep{roychowdhury2003effects}, the evanescent waves do not
contribute to the non-conservative force $F_{z}^{nc}\left(\mathbf{r},\omega\right)$.

In summary \textit{there are therefore two force components acting
on the particle: $F_{z,ev}^{grad}\left(z,\omega\right)$ and $F_{z,h}^{nc}\left(z,\omega\right)$,
perfectly distinguishable from each other since the former is due
to the non-radiative plane wave components of the emitted field, whereas
to the latter only the radiative components contribute. As the distance
from the particle to the source plane grows to values $z>\lambda$,
$F_{z,ev}^{grad}\left(z,\omega\right)$ tends to zero due to its evanescent
wave composition}. Nevertheless, as we shall see, the source coherence
length $\sigma$ plays an important role on these contributions.

The integration of Eqs. (\ref{Fgradhom}) and (\ref{Fncho}) using
the Gaussian spectral degree of coherence, quoted before: $\mu^{(0)}\left(\rho,\omega\right)=\textrm{exp}\left[-\rho^{2}/2\sigma^{2}\right]$,
leads to an analytical expression for the gradient and for the non-conservative
force. For the latter, Eq. (\ref{relfs}) yields the proportion of
radiation pressure and curl components for unpolarized emission. This
calculation is straightforwardly done on making: $s_{x}=s_{\perp}\cos\phi$,
$s_{y}=s_{\perp}\sin\phi$, and leads to 
\begin{eqnarray}
F_{z}^{grad}(z,\omega)=\Re\alpha_{e}S^{(0)}(\omega)e^{-\frac{\text{1}}{2}k^{2}\sigma^{2}}[\frac{z}{\sigma^{2}}\nonumber \\
-\sqrt{\frac{\pi}{2}}(\frac{2z^{2}}{\sigma^{3}}+\frac{1}{2\sigma})e^{\frac{2z^{2}}{\sigma^{2}}}\textrm{erfc}(\sqrt{2}z/\sigma)].\label{anal1}
\end{eqnarray}
 
\begin{eqnarray}
F_{z}^{nc}\left(z,\omega\right)=\Im\alpha_{e}S^{(0)}(\omega)/2[k\nonumber \\
-\frac{1}{\sigma}\sqrt{\frac{\pi}{2}}e^{-\frac{1}{2}k^{2}\sigma^{2}}\textrm{erfi}(k\sigma/\sqrt{2})],\label{anal2}
\end{eqnarray}
 where $\textrm{erfc}(x)=1-\textrm{erf}(x)$, $\textrm{erf}(x)$ being
the error function: $\textrm{erf}(x)=2/\sqrt{\pi}\int_{0}^{x}e^{-t^{2}}dt$,
and $\textrm{erfi}(x)$ is a positive real function defined as $\textrm{erfi}(x)=\textrm{erf}(ix)/i$.

\subsection{The {\it curl} force}
It is well-known \citep{Albala,wong2006gradient} that the non-conservative
part of the force $F_{i}^{nc}$ is the sum of a \textit{scattering
force}, or \textit{radiation pressure} 
\begin{eqnarray}
F_{i}^{nc}&=&(k/2)\Im\alpha_{e}\Re\left\langle \mathbf{E}\times\mathbf{B}^{*}\right\rangle _{i} \nonumber \\
&=&(1/2)\Im\alpha_{e}\Im\left\{ \left\langle E_{j}^{*}\partial_{i}E_{j}\right\rangle -\left\langle E_{j}^{*}\partial_{j}E_{i}\right\rangle \right\} ,
\end{eqnarray}
given by the averaged field Poynting vector, plus \textit{the curl
of a electric spin density}: 
\begin{eqnarray}
F_{i}^{nc,curl}&=&(1/2)\Im\alpha_{e}\Im\left\langle \left(\mathbf{E}^{*}\cdot\nabla\right)\mathbf{E}\right\rangle _{i} \nonumber \\
&=&(1/2)\Im\alpha_{e}\Im\left\langle E_{j}^{*}\partial_{j}E_{i}\right\rangle .
\end{eqnarray}

If the field emitted by the source is {\it unpolarized}: ${\cal E}_{jk}^{(0)}(\rho,\omega)=F^{(0)}(\rho,\omega)\delta_{jk}$,
$F^{(0)}(\rho,\omega)$ being a scalar spatial correlation function
whose two-dimensional Fourier transform will be denoted as $\tilde{F}^{(0)}(k{\bf s}_{\perp},\omega)$.
Then 
\begin{eqnarray}
\textrm{Tr}\mathcal{\tilde{E}}_{jk}^{(0)}(k\mathbf{s}_{\perp},\omega)
=3\tilde{F}^{(0)}(k\mathbf{s}_{\perp},\omega),
\end{eqnarray}
and the radiation pressure contribution $F_{i}^{nc,pr}$ to the non-conservative
force is: 
\begin{eqnarray}
&&F_{i}^{nc,pr}\nonumber\\
&&=\frac{k^{3}}{2}\Im\alpha_{e}\int_{\left|\mathbf{s}_{\perp}\right|\leq1}\left[\text{Tr}\tilde{{\cal E}}_{jk}\left(k\mathbf{s}_{\perp},\omega\right)s_{i}-\tilde{{\cal E}}_{ji}\left(k\mathbf{s}_{\perp},\omega\right)s_{j}\right]d^{2}\mathbf{s}_{\perp}\nonumber\\
&&=\frac{k^{3}}{2}\Im\alpha_{e}\int_{\left|\mathbf{s}_{\perp}\right|\leq1}\left[3\tilde{F}^{(0)}\left(k\mathbf{s}_{\perp}\right)-\tilde{F}^{(0)}\left(k\mathbf{s}_{\perp}\right)\right]s_{i}d^{2}\mathbf{s}_{\perp}\nonumber\\
&&=k^{3}\Im\alpha_{e}\int_{\left|\mathbf{s}_{\perp}\right|\leq1}\tilde{F}^{(0)}\left(k\mathbf{s}_{\perp}\right)s_{z}d^{2}\mathbf{s}_{\perp}=F_{z}^{nc,pr},
\end{eqnarray}
 since the azimuthal angle integrations when $s_{i}$ is either $s_{x}$
or $s_{y}$ is zero.

In a similar manner, the curl force contribution $F_{i}^{nc,curl}$
to $F_{i}^{nc}$ is 
\begin{eqnarray}
F_{i}^{nc,curl}&=&\frac{k^{3}}{2}\Im\alpha_{e}\int_{\left|\mathbf{s}_{\perp}\right|\leq1}\tilde{{\cal E}}_{ji}\left(k\mathbf{s}_{\perp},\omega\right)s_{j}d^{2}\mathbf{s}_{\perp}\nonumber\\
&=&\frac{k^{3}}{2}\Im\alpha_{e}\int_{\left|\mathbf{s}_{\perp}\right|\leq1}\tilde{F}^{(0)}\left(k\mathbf{s}_{\perp}\right)s_{z}d^{2}\mathbf{s}_{\perp}=F_{z}^{nc,curl}.\nonumber\\
&&
\end{eqnarray}

Namely, for unpolarizad radiation: 
\begin{equation}
F_{z}^{nc,pr}=2F_{z}^{nc,curl}=\frac{2}{3}F_{z}^{nc}.\label{relfs}
\end{equation}

\section{EXCITATION OF SURFACE PLASMON POLARITONS. NUMERICAL RESULTS}
Without loss of generality, we shall also address surface plasmon
polaritons (SPPs), excited on the source plane $z=0$. Let this be
gold for example, choosing for instance $\lambda=459.9nm$, its permittivity
is $\varepsilon=-2.546+i3.37$ \citep{palik1998handbook}. The SPP
wave vector $k\mathbf{s}_{\perp}=k\mathbf{s}_{\perp}^{SPP}=\pm k\left[\varepsilon/\left(\varepsilon+1\right)\right]^{1/2}$
corresponds to a pole of the Fresnel coefficient, (either on reflection
or on transmission depending on the set-up configuration), $R(k\mathbf{s}_{\perp},\omega)$
\citep{nietolibro,raether1988surface}. Then it is easy to obtain
that the former equations (\ref{Fgradhom}) and (\ref{Fnchye}) are
valid on substituting $\tilde{{\cal E}}_{jk}^{(0)}\left(k\mathbf{s}_{\perp},\omega\right)$
by $\tilde{{\cal E}}_{jk}^{(0)}\left(k\mathbf{s}_{\perp},\omega\right)\left|R\left(k\mathbf{s}_{\perp},\omega\right)\right|^{2}$
\citep{aunon2011near}

\begin{figure}
\begin{centering}
\includegraphics[scale=0.7]{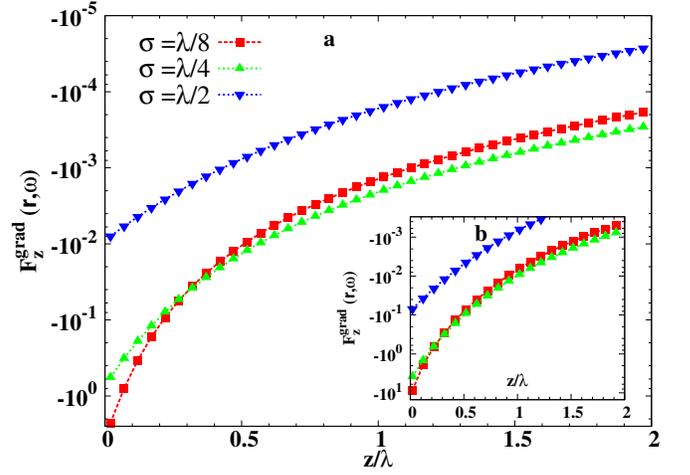} 
\par\end{centering}

\caption{Pulling gradient optical force due to evanescent components.
a, Gradient force in arbitrary units (a.u.) versus distance to the
source $z/\lambda$ for different values of the source coherence length
$\sigma$. b, The same force when SPPs are excited in the
source. A significant decrease of the magnitude of this force is clearly
seen as $\sigma$ grows about $\sigma=\lambda/2$.}
\end{figure}

Figure 2 shows the attractive gradient optical force due to evanescent
components for two random sources: one without and one with excited
SPPs (cf. Fig. 2a and Fig. 2b, respectively). The normalized value
$F_{z}^{grad}\left(z,\omega\right)/(kS^{(0)}\left(\omega\right)\Re\alpha_{e}/2\pi)$
is represented in arbitrary units. As predicted by Eq. (\ref{Fgradhom}),
the gradient force drags the particle towards the source plane; (notice
that since this normalization does not include $\Re\alpha_{e}$, it
does not contain an eventual negative value of this quantity). In
both figures we observe its exponential increase as the distance $z$
of the particle to the source decreases. \textit{Nevertheless, this
force is mainly governed by the coherence length} $\sigma$. For $\sigma=\lambda/8$
(red line), the magnitude of this force is maximum, but we observe
that around $\sigma=\lambda/2$ (blue line), and beyond, there is
an important decrease, with values between $10^{-2}$ and $10^{-3}$
in the magnitude even at subwavelength distances $z$ of this force,
which is practically zero, ($F_{z}^{grad}\simeq10^{-8}$), for $\sigma=\lambda$
and $z=0$ (this latter curve is not shown). With this, we demonstrate
that \textit{the decrease of the source coherence length gives rise
to an increase of the gradient force} and its effect is larger than
that of the distance $z$ of the particle to the source plane. Eventually,
a $\delta$-correlated source, (then $\sigma\rightarrow0$), like
e.g. a thermal source, will maximize this force. In addition, we show
with Fig. 2b that the excitation of SPPs in the source increases the
strength of this near field force by approximately one order of magnitude.
This is due to the then larger values of $\tilde{{\cal E}}_{jk}^{(0)}\left(k\mathbf{s}_{\perp},\omega\right)\left|R\left(k\mathbf{s}_{\perp},\omega\right)\right|^{2}$
stemming from the pole of $\left|R\left(k\mathbf{s}_{\perp},\omega\right)\right|^{2}$
at $k\mathbf{s}_{\perp}^{SPP}$.

\begin{figure}[h]
\begin{centering}
\includegraphics[scale=1.4]{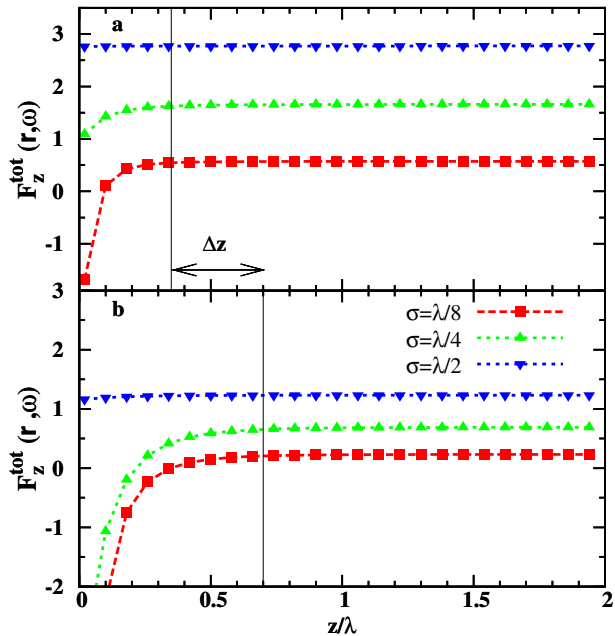} 
\par\end{centering}

\caption{Optical total force. a, Total force in arbitrary units (a.u.)
versus distance $z/\lambda$ to the source in arbitrary units for
different values of the coherence length $\sigma$. b, The
same as in (a) when SPPs are excited. In this second case we observe
an increment $\Delta z$ at which the magnitude of the gradient force
starts to exponentially increase}
\end{figure}

Correspondingly, Figures 3a and 3b show the normalized total force
$F_{z}^{tot}(z,\omega)=(2\pi/kS^{(0)}\left(\omega\right))\cdot(F_{z}^{grad}(z,\omega)/\Re\alpha_{e}+F_{z}^{nc}(z,\omega)/\Im\alpha_{e})$,
in arbitrary units, without and with SPP excitation, respectively.
At large distances $(z>\lambda)$, the total force is a constant of
the distance $z$ and repulsive according to the behaviour of the
non-conservative component $F_{z}^{nc}$, which dominates in this
region of $z$, regardless of the value of $\sigma$. In addition,
this non-conservative force is maximum for $\sigma=\lambda/2$, in
contrast with the decrease of the gradient component as $\sigma$
increases.

One might think that, due to its evanescent wave composition, the
magnitude of the gradient force at subwavelength distances would be
higher than that of the non-conservative force, however this is not
totally truth due the larger effect of the source coherence length
on $F_{z}^{grad}$ rather than on $F_{z}^{nc}$. In near-field $F^{tot}\simeq F^{grad}$
for $\sigma\leq\lambda/4$; but as $\sigma$ increases, $F^{grad}$
becomes negligible, being for $\sigma>\lambda/4$ $F^{tot}\simeq F^{nc}$.
These effects appear in Figs. 3a and 3b. Particulary, we see in
Fig. 3b that if SPPs are excited, an increment on the distance $\Delta z$
is produced where the gradient component cannot be neglected, (compare
Figs. 3a and 3b). The enhancement of the near field intensity due
to SPPs resonances then implies a longer-range of the this latter
pulling force.

\section{CONCLUSIONS}
We have reported a new area of study on photonic forces exerted on
small particles by discussing near field effects due to fluctuating
sources. This opens new perspectives on subwavelength effects and
manipulation in such general physics cases that range from light propagation
through the turbulent atmosphere \cite{radiophysics}, to speckle patterns from a large
variety of statistical sources, also including scatterers, optical
diffusers \cite{garcia97Transition, Riley00Three, Ripoll01Photon}, 
as well as thermal or blackbody sources, which opens new
possibilities at the subwavelength scale, particularly at the nanoscale.
We have seen that in the large variety of statistically stationary
and homogeneous sources, only the evanescent components contribute
to the gradient forces, while the non-conservative part that contains
radiation pressure and curl forces is due solely to emitted propagating
components. Hence the subwavelength information is encoded in the
gradient forces. Same numerical examples were given for statistically
isotropic unpolarized emitted wavefields, showing the important effect
that the source coherence length has on these forces, specially on
the gradient component. Also the excitation of surface waves importantly
enhances the magnitude of these forces. We expect that these findings
stimulate experiments and applications by this novel particle manipulation
scenario.

\section*{Acknowledgements}

The authors acknowledge support from the Spanish Ministerio de Ciencia
e Innovacin (MICINN) through the Consolider NanoLight CSD2007-00046
and FIS2009-13430-C02-01 research grants. J. M. Au\~{n}\'{o}n thanks a scholarship
from MICINN.


%

\end{document}